\pdfoutput=1
\documentclass[useAMS,usenatbib]{mn2e}
\usepackage{apjfonts}
\usepackage{graphicx}
\usepackage{amsmath}
\usepackage{amssymb}
\usepackage{ctable}
\usepackage{multirow}
\usepackage{fixltx2e}
\usepackage{threeparttable}
\usepackage{hyperref}
\hypersetup{colorlinks=true,linkcolor=blue,citecolor=blue,filecolor=blue,urlcolor=blue}

\newcommand{\be}{\begin{equation}}
\newcommand{\ee}{\end{equation}}
\newcommand{\ba}{\begin{eqnarray}}
\newcommand{\ea}{\end{eqnarray}}

\newcommand{\Mhalo}{M_{\rm halo}}

\newcommand{\Ms}{M_{\ast}}

\newcommand{\cm}{{\rm cm}}

\newcommand{\pc}{{\rm pc}}
\newcommand{\kpc}{{\rm kpc}}
\newcommand{\Mpc}{{\rm Mpc}}

\newcommand{\Myr}{{\rm Myr}}

\newcommand{\Msun}{M_{\sun}}

\newcommand{\mb}{m_b}
\newcommand{\mDM}{m_{\rm DM}}
\newcommand{\eg}{\epsilon_{\rm gas}}
\newcommand{\es}{\epsilon_{\rm star}}
\newcommand{\eDM}{\epsilon_{\rm DM}}

\newcommand{\nc}{n_{\rm th}}

\newcommand{\Mc}{M_{\rm cl}}

\newcommand{\fesc}{f_{\rm esc}}

\newcommand{\eff}{\epsilon_{\rm ff}}
\newcommand{\dd}{{\rm d}}
\newcommand{\LCDM}{$\Lambda$CDM}
\newcommand{\mol}{{\sc mol}}
\newcommand{\sg}{{\sc sg}}
\newcommand{\den}{{\sc den}}
\newcommand{\cf}{{\sc cf}}

\newcommand{\apj}{ApJ}
\newcommand{\apjl}{ApJ}
\newcommand{\apjs}{ApJS}
\newcommand{\mnras}{MNRAS}
\newcommand{\aap}{A\&A}
\newcommand{\araa}{ARA\&A}

\newcommand{\pasa}{PASA}
\newcommand{\aapr}{A\&ARv}
\newcommand{\rmxaa}{RMxAA}

\def\femcl{f_{\rm emitted}^{\rm cl}}
\def\fesccl{f_{\rm escaped}^{\rm cl}}

\title[$\fesc$ of Proto-GCs]
{The contribution of globular clusters to cosmic reionization}

\author[X. Ma et al.]{
  \parbox[t]{1.0\textwidth}{
   Xiangcheng Ma,$^{1}$\thanks{E-mail: \href{mailto:xchma@berkeley.edu}{xchma@berkeley.edu}}
   Eliot Quataert,$^1$
   Andrew Wetzel,$^2$ \\
   Claude-Andr{\'e} Faucher-Gigu{\`e}re$^3$ and
   Michael Boylan-Kolchin$^4$
  }
  \vspace{5pt} \\
  $^1$Department of Astronomy and Theoretical Astrophysics Center, University of California Berkeley, Berkeley, CA 94720 \\
  $^2$Department of Physics, University of California, Davis, CA 95616, USA \\
  $^3$Department of Physics and Astronomy and CIERA, Northwestern University, 2145 Sheridan Road, Evanston, IL 60208, USA \\
  $^4$Department of Astronomy, The University of Texas at Austin, 2515 Speedway Blvd, Stop C1400, Austin, TX 78712, USA \\
}

\pagerange{\pageref{firstpage}--\pageref{lastpage}}
\date{Draft version \today}

\begin{document}
\maketitle
\label{firstpage}

\begin{abstract}
We study the escape fraction of ionizing photons ($\fesc$) in two cosmological zoom-in simulations of galaxies in the reionization era with halo mass $\Mhalo\sim10^{10}$ and $10^{11}\,\Msun$ (stellar mass $\Ms\sim10^7$ and $10^9\,\Msun$) at $z=5$ from the Feedback in Realistic Environments project. These simulations explicitly resolve the formation of proto-globular clusters (GCs) self-consistently, where 17--39\% of stars form in bound clusters during starbursts. Using post-processing Monte Carlo radiative transfer calculations of ionizing radiation, we compute $\fesc$ from cluster stars and non-cluster stars formed during a starburst over $\sim100$\,Myr in each galaxy. We find that the averaged $\fesc$ over the lifetime of a star particle follows a similar distribution for cluster stars and non-cluster stars. Clusters tend to have low $\fesc$ in the first few Myrs, presumably because they form preferentially in more extreme environments with high optical depths; the $\fesc$ increases later as feedback starts to destroy the natal cloud. On the other hand, some non-cluster stars formed between cluster complexes or in the compressed shells at the front of a superbubble can also have high $\fesc$. We find that cluster stars on average have comparable $\fesc$ to non-cluster stars. This result is robust across several star formation models in our simulations. Our results suggest that the fraction of ionizing photons from proto-GCs to cosmic reionization is comparable to the cluster formation efficiencies in high-redshift galaxies and thus proto-GCs likely contribute an appreciable fraction of photons but are not the dominant sources for reionization.
\end{abstract}

\begin{keywords}
galaxies: evolution -- galaxies: formation -- galaxies: high-redshift -- cosmology:
theory -- globular clusters: general -- dark ages, reionization, first stars
\end{keywords}

\section{Introduction}
\label{sec:intro}
Globular clusters (GCs) are the fossil record of some of the earliest star formation in the Universe: they are compact, tightly bound stellar systems that are usually old ($\rm age\sim5$--13\,Gyr) and metal-poor ($-2.5\lesssim\rm [Fe/H]\lesssim0$) with half-light radii 0.5--10\,pc (see e.g. \citealt{Harris:1991,Brodie:2006,Kruijssen:2014,Bastian:2018,Gratton:2019,Krumholz:2019}, for a series of reviews). The idea that GCs form in high-pressure regions of the interstellar medium (ISM) during regular star formation at high redshifts \citep[e.g.][]{Elmegreen:1997,Kruijssen:2012b} has been adopted in most recent models of GC formation in a cosmological context \citep[e.g.][]{Kravtsov:2005,Li:2014,Li:2017,Choksi:2018,Pfeffer:2018,Kruijssen:2019,El-Badry:2019}, although other scenarios have been proposed to explain the formation of GCs \citep[e.g.][]{Peebles:1968,Peebles:1984,Fall:1985,Naoz:2014,Kimm:2016,Madau:2020}.

It has long been speculated that the progenitors of present-day GCs formed in high-redshift galaxies likely contribute a significant fraction of ionizing photons to reionize the Universe. For example, \citet{Ricotti:2002} presented a model where the proto-GCs can provide all the ionizing photons required for reionization if they all formed at $z\gtrsim5$ \citep[see also e.g.][]{Schaerer:2011,Katz:2013,Katz:2014,Boylan-Kolchin:2018} under two assumptions: (a) the progenitors are $\sim10$ times more massive than survived GCs to account for mass loss due to dynamical evolution, and (b) proto-GCs have ionizing photon escape fraction $\fesc\sim1$.\footnote{In this paper, we refer to $\fesc$ as the fraction of photons that escape the halo virial radius, unless stated otherwise. We will study instantaneous and lifetime-averaged $\fesc$ for individual stars and galaxy-averaged $\fesc$.} The first assumption implies a large fraction of stars in high-redshift galaxies formed in proto-GCs. This is possibly true given the highly gas-rich, turbulent ISM in $z\gtrsim5$ galaxies and can be checked in GC formation models that match the observed GC population in the local Universe \citep[e.g.][]{Katz:2013,Katz:2014,Choksi:2018,Reina-Campos:2018,Reina-Campos:2019,Pfeffer:2019}.

The second assumption above has not been fully investigated. Recently, \citet{He:2020} studied the $\fesc$ from star-forming clouds using a suite of isolated molecular cloud simulations with different cloud mass and sizes that include detailed treatments of single star formation and photoionization feedback. They found that $\fesc$ {\em from the cloud} increases with increasing cloud compactness at the same mass \citep[see also][]{Howard:2017,Kim:2019}. This suggests that proto-GCs, presumably formed in more compact clouds, likely show higher $\fesc$ than stars formed in less extreme conditions. It is, however, unclear if such isolated clouds represent all star formation environments in high-redshift galaxies. For example, galactic scale simulations point out the importance of gas accretion from sub-kpc scales during star cluster formation \citep[see e.g.][]{Lahen:2020,Ma:2020a}; observations and simulations suggest that star formation may also be triggered by stellar feedback \citep[see e.g.][]{Heiles:1979,Pellegrini:2012,Ma:2020}. Moreover, these studies do not account for the galactic ISM and the gaseous halo that are critical for understanding the $\fesc$ to the intergalactic medium \citep[which is directly relevant to cosmic reionization; e.g.][]{Kuhlen:2012,Robertson:2013,Robertson:2015,Finkelstein:2019,Faucher-Giguere:2020}.

Cosmological hydrodynamic simulations of galaxy formation have been a powerful tool for predicting $\fesc$ from $z\gtrsim5$ galaxies \citep[e.g.][]{Yajima:2011,Wise:2014,Kimm:2014, Ma:2015,Ma:2016,Ma:2020,Paardekooper:2015,Xu:2016b,Rosdahl:2018}. It has recently become possible to explicitly resolve the formation of proto-GCs in state-of-the-art simulations \citep[see e.g.][]{Kim:2018,Mandelker:2018,Lahen:2020,Ma:2020a}. In this paper, we utilize a suite of cosmological zoom-in simulations from the Feedback in Realistic Environments (FIRE)\footnote{\url{https://fire.northwestern.edu}} project with self-consistently resolved proto-GC formation in galaxies at $z\gtrsim5$ \citep{Ma:2020a} and post-processing Monte Carlo radiative transfer (MCRT) calculations to study the $\fesc$ from proto-GCs in a realistic galactic and cosmological environment. We also compare $\fesc$ from clusters and non-cluster stars to understand the relative contribution of proto-GCs to cosmic reionization.

In recent years, a few highly magnified sources with compact sizes (effective radii 10--100\,pc) at $z\sim2$--6 have been discovered in strong gravitational lensing fields of galaxy clusters. These sources are bright with extremely high star formation rate surface densities \citep[e.g.][]{Vanzella:2017,Vanzella:2017a,Vanzella:2019,Zick:2020}, which are believed to be actively forming proto-GCs, super star clusters, or star cluster complexes at high redshifts. Among these sources, the {\em Sunburst} arc \citep{Rivera-Thorsen:2017}, a strongly lensed arc at $z=2.37$ that contains a compact star-forming knot likely to be a proto-GC, has been confirmed to leak ionizing radiation \citep{Rivera-Thorsen:2019}. This is empirical evidence that proto-GCs may play a significant role in reionization \citep[see also e.g.][]{Vanzella:2020}. Our results will be useful for understanding these observations and making plans for future programs using the upcoming {\em James Webb Space Telescope} (JWST).

The paper is organized as follows. In Section \ref{sec:method}, we review the simulations studied in this work and the method used for our analysis. We present the $\fesc$ from proto-GCs in our simulations in Section \ref{sec:fesc} and provide some physical insights in Section \ref{sec:physics}. We discuss our results and conclude in Section \ref{sec:discussion}. 

We adopt a standard flat {\LCDM} cosmology with {\it Planck} 2015 cosmological parameters $H_0=68 {\rm\,km\,s^{-1}\,Mpc^{-1}}$, $\Omega_{\Lambda}=0.69$, $\Omega_{m}=1-\Omega_{\Lambda}=0.31$, $\Omega_b=0.048$, $\sigma_8=0.82$, and $n=0.97$ \citep{Planck-Collaboration:2016a}. We use a \citet{Kroupa:2002} initial mass function (IMF) from 0.1--$100\,\Msun$, with IMF slopes of $-1.30$ from 0.1--$0.5\,\Msun$ and $-2.35$ from 0.5--$100\,\Msun$.

\begin{table}
\caption{Cosmological zoom-in simulations studied in this paper.}
\begin{threeparttable}
\begin{tabular}{ccccccc}
\hline
Name & $\Mhalo$ & $\Ms$ & $\mb$ & $\eg$ & $\mDM$ & $\eDM$ \rule{0pt}{8pt} \\
 & [$\Msun$] & [$\Msun$] & [$\Msun$] & [pc] & [$\Msun$] & [pc] \rule[-5pt]{0pt}{0pt} \\
\hline
z5m10a & $6.6\times10^9$ & $1.5\times10^7$ & 119.3 & 0.14 & 651.2 & 10 \rule{0pt}{9pt} \\
z5m11c & $7.6\times10^{10}$ & $9.4\times10^8$ & 890.8 & 0.28 & 4862.3 & 21 \rule[-4pt]{0pt}{0pt} \\
\hline
\end{tabular}
\begin{tablenotes}
\item {\em Notes.} (1) $\Mhalo$ and $\Ms$: Halo mass and total stellar mass inside the halo virial radius at $z=5$. (2) $\mb$ and $\mDM$: Initial masses of baryonic and dark matter particles in the zoom-in regions. (3) $\eg$ and $\eDM$: Plummer-equivalent force softening lengths for gas and dark matter particles, in comoving units at $z\geq9$ and physical units at $z<9$. Force softening for gas is adaptive ($\eg$ is the minimum softening length). Force softening length for star particles is $\es=5\eg$.
\end{tablenotes}
\end{threeparttable}
\label{tbl:sim}
\end{table} 

\begin{figure*}
\centering
\includegraphics[width=0.89\linewidth]{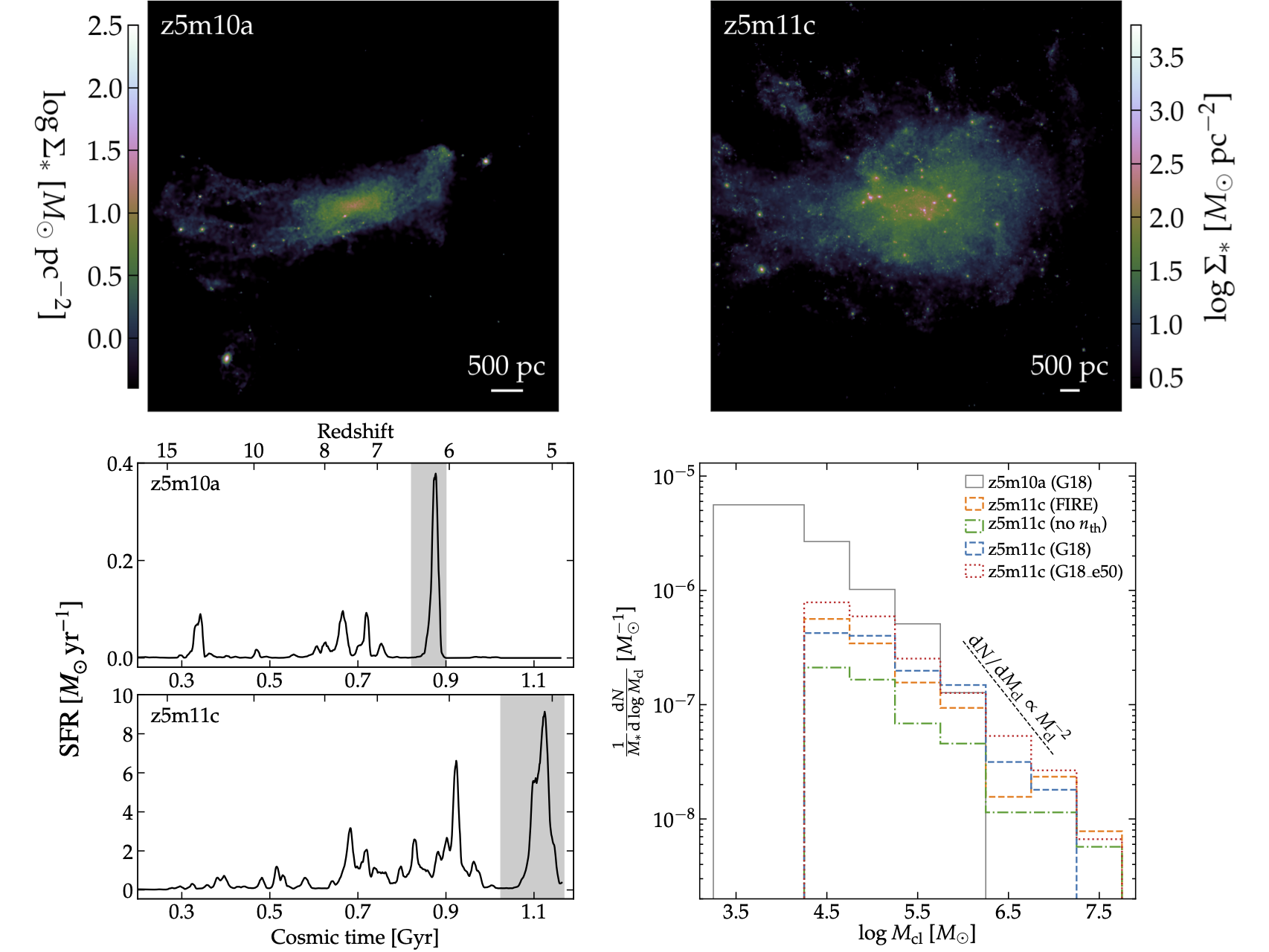}
\caption{{\it Top}: Stellar images of cosmological zoom-in simulations z5m10a (left) and z5m11c (right) at their final redshift ($z=5$). {\it Bottom left}: Star formation histories of the two simulated galaxies. The grey shaded regions label the starburst we identified and re-simulated in each simulation (taken from figs. 1 and 2 in \citealt{Ma:2020a}). {\it Bottom right}: Normalized mass function of bound star clusters formed in the re-simulations studied in this paper (see Section \ref{sec:sims} and Table \ref{tbl:resim} for details). The black dashed line shows the power-law function $\dd N/\dd \Mc \propto \Mc^{-2}$ (reproduced from figs. 10 and 16 in \citealt{Ma:2020a}).}
\label{fig:review} 
\end{figure*}

\begin{table*}
\caption{Simulation restarts. From each parent simulation, a starburst is re-started and run for the cosmic time and redshift labeled (Fig. \ref{fig:review}) with different star formation models. We consider the following criteria for star formation: (i) molecular (\mol), (ii) self-gravitating (\sg), (iii) density threshold (\den), and (iv) converging flow (\cf) (see descriptions in Sections \ref{sec:sims}). Each model refers to a combination of star formation criteria with certain choice of local star formation efficiency $\eff$ listed below.}
\begin{threeparttable}
\begin{tabular}{ccccccccc}
\hline
Parent & Redshift & Cosmic time & $\Mhalo$ & Model & Star formation & $\eff$ & $\Ms$ formed & $f_{\rm bound}$ \rule{0pt}{8pt} \\
simulation & range & [Gyr] & [$\Msun$] & name & criteria & & [$\Msun$] & \rule[-5pt]{0pt}{0pt} \\
\hline
z5m10a & 6.605--6.143 & 0.820--0.901 & $5.8\times10^9$ & G18 & {\mol+\sg+\cf} & 1 & $1.57\times10^7$ & 0.26 \rule{0pt}{9pt} \\
z5m11c & 5.562--5.024 & 1.023--1.163 & $7.6\times10^{10}$ & FIRE & {\mol+\sg+\den} & 1 & $2.56\times10^8$ & 0.28  \\
 & & & & no $\nc$ & {\mol+\sg} & 1 & $3.50\times10^8$ & 0.17 \\
 & & & & G18 & {\mol+\sg+\cf} & 1 & $4.44\times10^8$ & 0.26 \\
 & & & & G18\_e50 & {\mol+\sg+\cf} & 0.5 & $3.00\times10^8$ & 0.39 \rule[-4pt]{0pt}{0pt} \\
\hline
\end{tabular}
\begin{tablenotes}
\item {\em Notes.} (1) Parent simulation: The cosmological zoom-in simulation where the starburst is selected (see Table \ref{tbl:sim} and Fig. \ref{fig:review}). (2) Redshift and cosmic time: The redshift and cosmic time interval where the starburst has been re-simulated. (3) $\Mhalo$: Average halo mass during the re-simulation. (4) $\Ms$ formed: Total stellar mass formed in the halo during the burst. (5) $f_{\rm bound}$: The fraction of stars formed during the re-simulation that end up in bound clusters by the end of the re-simulation.
\end{tablenotes}
\end{threeparttable}
\label{tbl:resim}
\end{table*}

\section{Methods}
\label{sec:method}
\subsection{The simulations} 
\label{sec:sims}
The simulations we analyze in this paper were first presented in \citet{Ma:2020a}. We only focus on two cosmological zoom-in simulations, z5m10a and z5m11c, selected from a cosmological volume of $(30\,h^{-1}\,\Mpc)^3$ with approximate halo mass $10^{10}$ and $10^{11}\,\Msun$ at $z=5$, respectively. These simulations were first run to $z=5$ using {\sc gizmo}\footnote{\url{http://www.tapir.caltech.edu/~phopkins/Site/GIZMO.html}} \citep{Hopkins:2015} in its mesh-less finite-mass (MFM) mode and default FIRE-2 models of the multi-phase ISM, star formation, and feedback from \citet{Hopkins:2018b}. We have saved 67 snapshots from $z=20$--5 with time interval $\sim16$--20\,Myr between successive snapshots. The final halo mass and stellar mass by $z=5$ in both simulations, along with the mass resolution\footnote{Note that the two simulations were named z5m10a\_hr and z5m11c\_hr in \citet{Ma:2020a} to differentiate them from previous runs of the same halo with 8 times lower mass resolution. We only analyze the runs with the best resolution in this work.} and force softening lengths adopted in these runs are provided in Table \ref{tbl:sim}. In Fig.~\ref{fig:review}, we show the stellar image at $z=5$ (top row) and star formation history (bottom left) for these simulations. We identified a starburst in each simulation and re-simulated the burst from a snapshot prior to the burst for about $\sim100$\,Myr using various star formation prescriptions different from the default FIRE-2 model (see descriptions below and in Table \ref{tbl:resim} for details). We saved a snapshot every 0.5\,Myr in the re-simulations to track the formation process of GC candidates. The shaded regions in the bottom-left panel of Fig. \ref{fig:review} label the starbursts we re-simulated. The burst in z5m10a is caused by merger, while the one in z5m11c is triggered by rapid gas accretion onto the ISM (no merger happened at this epoch).

We briefly review the baryonic physics included in our simulations below, but refer to \cite{Hopkins:2018b} for more details on the numerical implementations and tests. Gas follows an ionized-atomic-molecular cooling curve between 10 and $10^{10}$\,K, including metallicity-dependent fine-structure and molecular cooling at low temperatures and metal-line cooling at high temperatures. The ionization states and cooling rates for H and He are computed following \cite{Katz:1996} and cooling rates for heavy elements are calculated from a compilation of {\sc cloudy} runs \citep{Ferland:2013}, applying a uniform, redshift-dependent ionizing background from \cite{Faucher-Giguere:2009} and heating from local sources. Self-shielding is accounted for with a local Sobolev approximation.

We consider combinations of the following star formation criteria: (i) {\em Molecular} (\mol). We estimate the self-shielded molecular fraction for each gas particle ($f_{\rm mol}$) following \cite{Krumholz:2011}. Stars only form in molecular gas.

(ii) {\em Self-gravitating} (\sg). Star formation is allowed only when the gravitational potential energy is larger than kinetic plus thermal energy at the resolution scale, described by the virial parameter
\be
\label{eqn:virial}
\alpha \equiv \frac{\| \nabla\otimes\mathbf{v} \|^2_i + (c_{{\rm s},\,i}/h_i)^2}{8\pi G \, \rho_i} < 1,
\ee
where $\otimes$ is the outer product, $c_{\rm s}$ is the sound speed, $h$ is the resolution scale, and the subscript $i$ implies that the quantities are evaluated for individual gas particles \citep{Hopkins:2013b}.

(iii) {\em Density threshold} (\den). The number density of hydrogen exceeds a threshold of $n_{\rm H}>\nc=10^3\,\cm^{-3}$. 

(iv) {\em Converging flow} (\cf). Star formation is restricted to converging flows where $\nabla\cdot{\bf v}<0$.

The default FIRE-2 model for star formation consists of criteria \mol, \sg, and \den. In the re-simulations, we also consider two alternative models: `no $\nc$' ({\mol} and {\sg}) and `G18' (\mol, \sg, and \cf; \citealt{Grudic:2018}). If the criteria above are met, a gas particle will turn into a star particle at a rate $\dot{\rho}_{\ast}=\eff \, f_{\rm mol} \,\rho/t_{\rm ff}$, where $\eff$ is the {\em local} star formation efficiency and $t_{\rm ff}$ is the free-fall time at the density of the particle. We adopt $\eff=1$ by default. In addition, we consider another model `G18\_e50' which uses the `G18' criteria and $\eff=0.5$. We reiterate that $\eff$ represents the rate where locally self-gravitating clumps fragment, while the cloud-scale star formation efficiencies are regulated by feedback at $\sim1$--10\% per {\em cloud} free-fall time for typical conditions of Milky Way molecular clouds (see \citealt{Hopkins:2018b} and references therein for extensive tests). Table \ref{tbl:resim} lists all re-simulations we analyze in this paper. We refer to \cite{Ma:2020a} for detailed comparison among these models and discussion on the differences.

Every star particle is treated as a single stellar population with known age, mass, and metallicity (inherited from its parent gas particle). All feedback quantities are calculated directly from standard stellar population synthesis models {\sc starburst99} \citep{Leitherer:1999} assuming a \cite{Kroupa:2002} IMF. The simulations account for the following feedback mechanisms: (i) photoionization and photo- electric heating, (ii) local and long-range radiation pressure for UV and optical single scattering and multiple scattering of infrared re-radiated photons, and (iii) energy, momentum, mass, and metal return from discrete supernovae (SNe) and continuous stellar winds (OB/AGB stars). More details on the numerical implementations of these feedback channels are presented in \cite{Hopkins:2018b}. We note that FIRE uses a very approximate model for photoionization from stars, which does not make significant differences on galaxy-scale dynamics compared to radiation-hydrodynamic method \citep[e.g.][]{Hopkins:2020}, but we rely on post-processing MCRT calculations to obtain $\fesc$ (Section \ref{sec:mcrt}; for more detailed discussion, see \citealt{Ma:2020}, section 2.1). We include metal yields from Type-II SNe, Type-Ia SNe, and AGB winds and adopt a sub-grid turbulent metal diffusion and mixing algorithm described in \citet{Su:2017} and \citet{Escala:2018}.

\begin{figure}
\centering
\includegraphics[width=\linewidth]{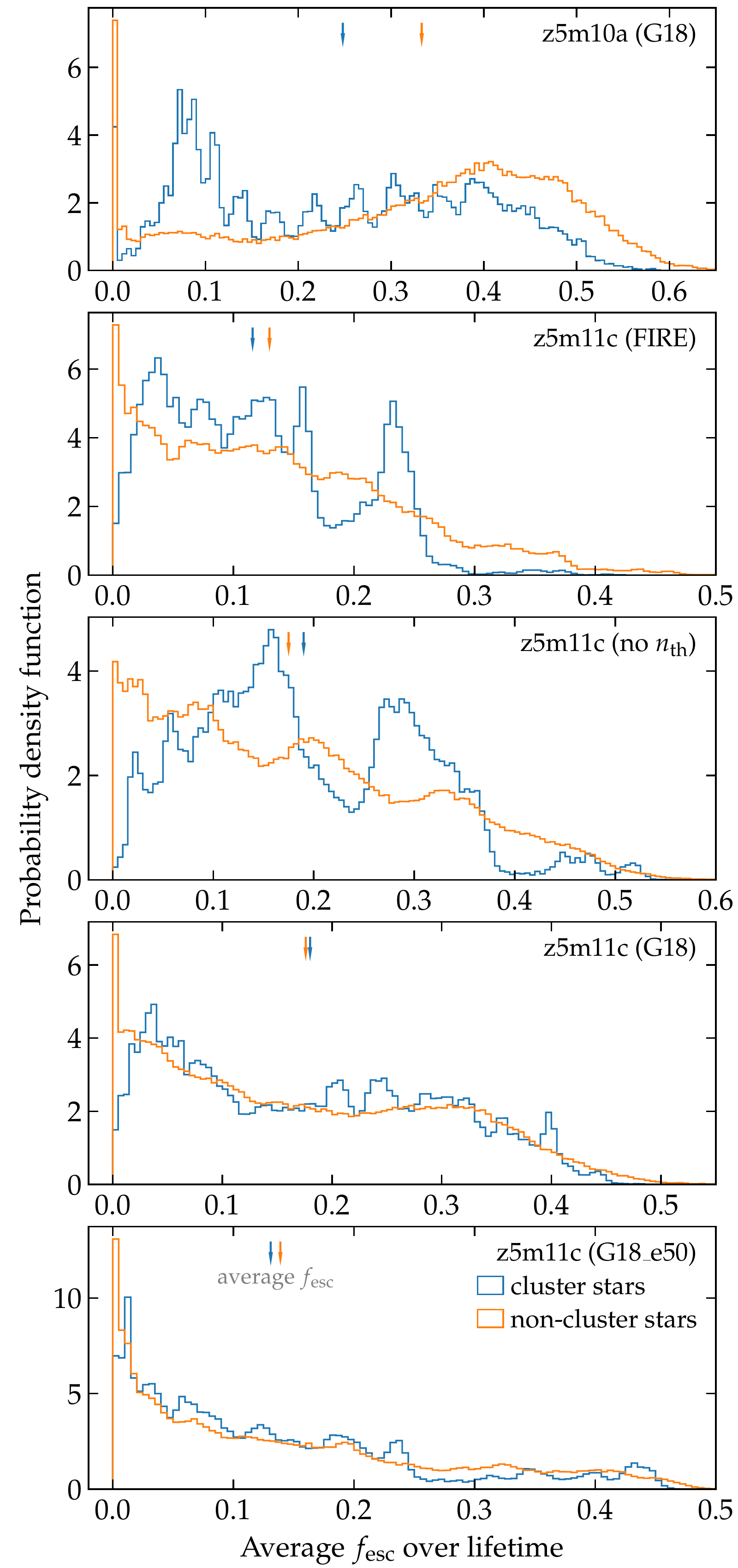}
\caption{Probability density functions of lifetime-averaged $\fesc$ for cluster stars (blue) and non-cluster stars (orange) in each re-simulation. Aside from an enhanced probability of $\fesc\sim0$ for non-cluster stars, there is no large difference in the distribution of $\fesc$ between the two groups. Arrows show the average $\fesc$ over all cluster and non-cluster stars, which are comparable in most runs (see Table \ref{tbl:fesc}).}
\label{fig:fdist} 
\end{figure}

\begin{table*}
\caption{The contribution of ionizing photons from proto-GCs in each re-simulation. {\em Middle}: The fraction of ionizing photons from bound star clusters, both for emitted ($\femcl$; left) and escaped ($\fesccl$; right) photons. {\em Right}: The average $\fesc$ over all stars (left), only cluster stars (center), and only non-cluster stars (right). We do not find significant difference on the average $\fesc$ among these groups.}
\begin{threeparttable}
\begin{tabular}{cc|p{1.5cm}p{1.5cm}|p{1.5cm}p{1.5cm}p{1.5cm}}
\hline 
\multirow{4}{*}{\parbox[t]{1.2cm}{\centering Parent simulation}} & \multirow{4}{*}{Model} & \multicolumn{2}{c|}{Fraction of ionizing photons} & \multicolumn{3}{c}{\multirow{2}{*}{Average escape fractions ($\fesc$)}} \\
 & & \multicolumn{2}{c|}{from bound clusters} & & & \\ \cline{3-7}
 & & \hfil Emitted & \hfil Escaped & \hfil All & \hfil Cluster & \hfil Non-cluster \rule{0pt}{8pt} \\ 
 & & \hfil ($\femcl$) & \hfil ($\fesccl$) & \hfil stars & \hfil stars & \hfil stars \rule[-4pt]{0pt}{0pt} \\
\hline
z5m10a & G18 & \hfil 0.235 & \hfil 0.186 & \hfil 0.313 & \hfil 0.248 & \hfil 0.333 \\
z5m11c & FIRE & \hfil 0.271 & \hfil 0.249 & \hfil 0.126 & \hfil 0.116 & \hfil 0.130 \\
z5m11c & no $n_{\rm th}$ & \hfil 0.176 & \hfil 0.189 & \hfil 0.178 & \hfil 0.190 & \hfil 0.175 \\
z5m11c & G18 & \hfil 0.261 & \hfil 0.266 & \hfil 0.177 & \hfil 0.180 & \hfil 0.176 \\
z5m11c & G18\_e50 & \hfil 0.399 & \hfil 0.385 & \hfil 0.136 & \hfil 0.131 & \hfil 0.139 \\
\hline
\end{tabular}
\end{threeparttable}
\label{tbl:fesc}
\end{table*} 

\subsection{Proto-GCs formed in the re-simulations}
\label{sec:gc}
One important advantage of our simulations is that they are able to explicitly resolve the formation of clusters in a self-consistent way. In \citet{Ma:2020a}, we studied the proto-GC populations formed during the re-simulated starbursts. Our findings are briefly summarized below. We run the cluster finder from \citet{Grudic:2018} on all star particles in the final snapshot of each re-simulation to identify gravitationally bound star clusters formed during the starburst. In the bottom-right panel of Fig. \ref{fig:review}, we show the cluster mass functions for all re-simulations analyzed in this paper, normalized by the total stellar mass formed in the re-simulation (reproduced from figs. 10 and 16 in \citealt{Ma:2020a}). The newly formed star clusters broadly follow the power-law mass function $\dd N/\dd \Mc \propto \Mc^{-2}$ (black dashed line). The formation efficiency of clusters at a fixed mass depends on the star formation model adopted in our simulations: a stricter (looser) model leads to a factor of a few more (less) clusters per unit stellar mass formed in the galaxy (see \citealt{Ma:2020a}, section 5 for more discussion). About 17--39\% of the stars formed in our re-simulated bursts belong to a cluster at the end of the re-simulations (see $f_{\rm bound}$ in Table \ref{tbl:resim}, last column).

Most of the clusters are compact with half-mass radii 6--40\,pc and have small metallicity spread ($\sigma_{\rm [Z/H]}\sim0.08$\,dex). The clusters are preferentially formed in high-pressure regions with gas surface densities $\sim10^4\,\Msun\,\pc^{-2}$, normally created by compression due to cloud-cloud collision or feedback-driven winds. The time-scales of cluster formation are short, from less than 2\,Myr for clusters below $10^5\,\Msun$ to $\sim5$\,Myr for clusters above $10^{6.5}\,\Msun$ in z5m11c. These clusters likely represent the progenitors of present-day GCs at high redshift. We refer to these gravitationally bound clusters formed in the re-simulations as `star clusters', or `proto-GCs' interchangeably in the rest of this paper. We do not consider short-lived clusters that are already disrupted at the end of the re-simulations, but we have checked this has little effect on our conclusions in this paper.

\begin{figure}
\centering
\includegraphics[width=\linewidth]{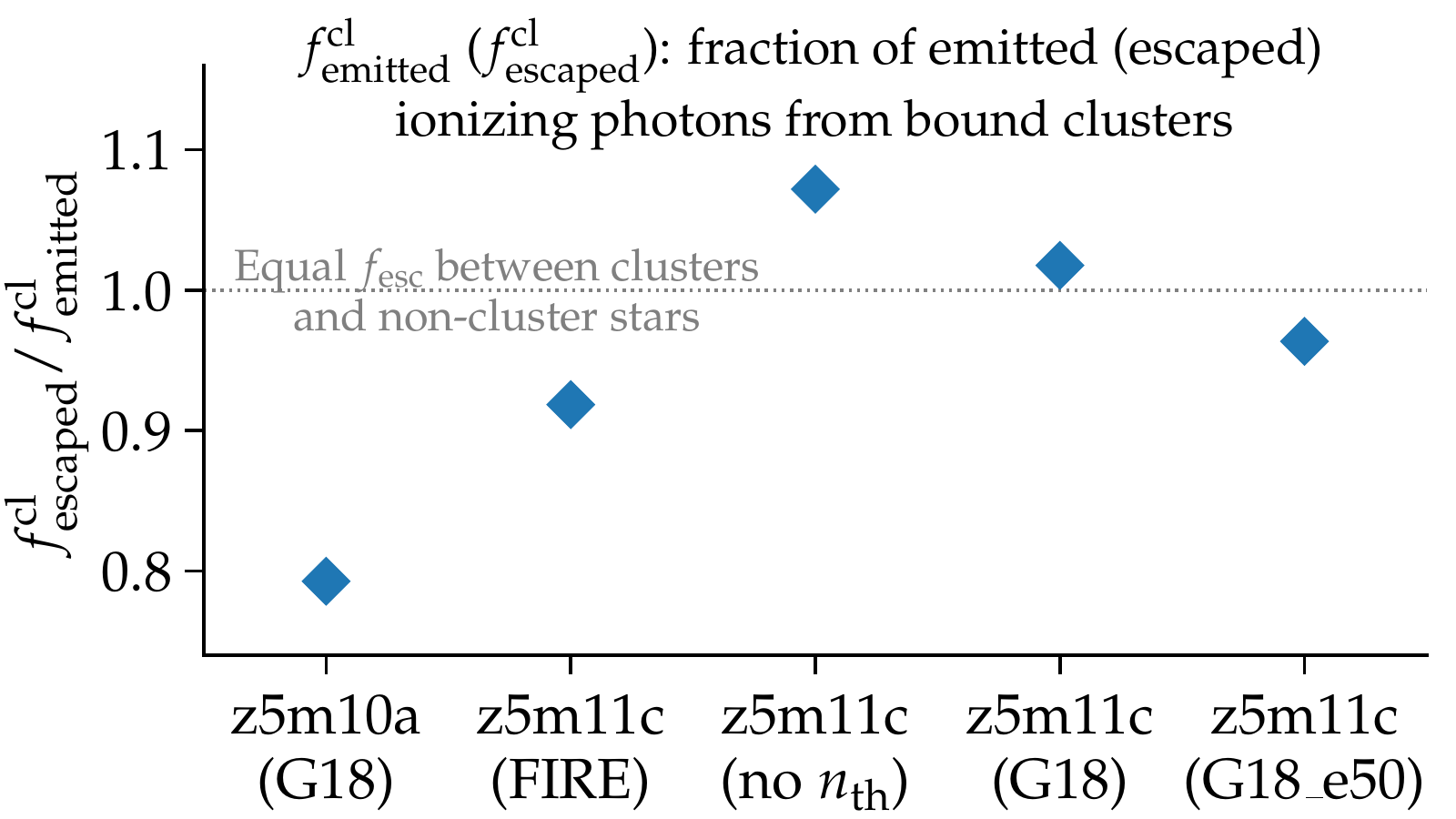}
\caption{The ratio of the fraction of escaped ionizing photons from bound clusters ($\fesccl$) to that of emitted photons ($\femcl$) for each re-simulation (both $\femcl$ and $\fesccl$ are listed in Table \ref{tbl:fesc}). Proto-GCs have similar $\fesc$ to other stars.}
\label{fig:frac} 
\end{figure}

\begin{figure}
\centering
\includegraphics[width=\linewidth]{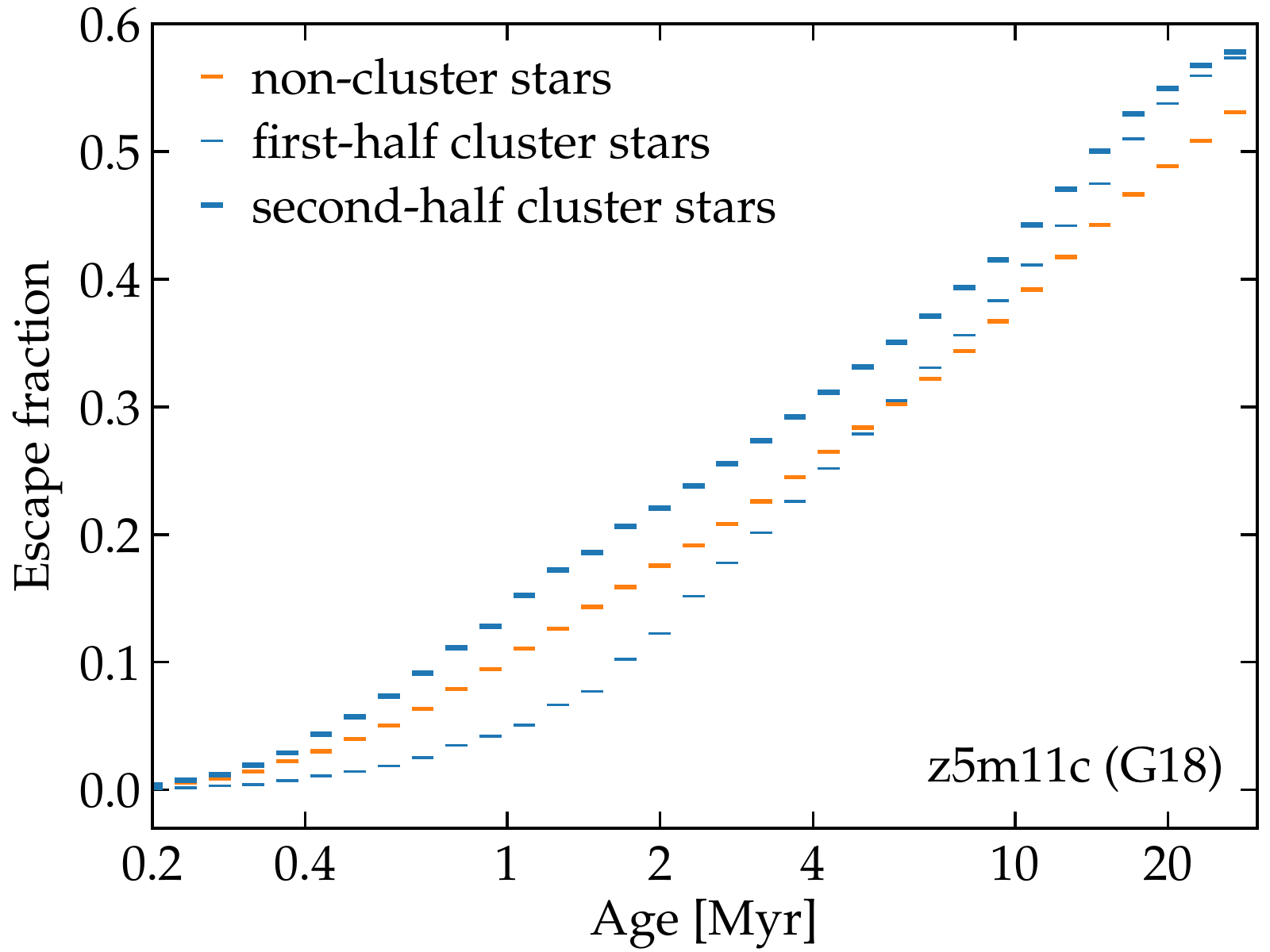}
\caption{The average $\fesc$ over stars in narrow stellar age bins in z5m11c (G18), divided into three groups: non-cluster stars (orange), the first half of stars formed in a cluster (thin blue), and the second half of stars formed in a cluster (thick blue). The first-half cluster stars have lower $\fesc$ than non-cluster stars in the first 3\,Myr, likely because they preferentially formed in high-pressure regions where the optical depths are high. The second-half of stars formed in clusters have higher $\fesc$ as feedback starts to disrupt the natal cloud. On average, though, cluster stars have comparable $\fesc$ to that of non-cluster stars.}
\label{fig:fage} 
\end{figure}

\subsection{The MCRT calculations}
\label{sec:mcrt}
For each snapshot saved in these re-simulations, we post-process it with a MCRT code of ionizing radiation to calculate $\fesc$ following the same approach as in \cite{Ma:2020}. We first map all the gas particles inside the halo virial radius onto an octree grid, where we adaptively refine the dense regions until no cell contains more than two gas particles. We emit $3.6\times10^8$ photon packets from star particles in the grid sampled by their ionizing photon emissivities and another $3.6\times10^8$ packets from the boundary of the grid inwards to represent the uniform, metagalactic ionizing background following the intensity from \citet{Faucher-Giguere:2009}. The photon packets are propagated in the grid until they are absorbed at some place or escape the domain. We account for absorption and scattering by neutral gas and dust grains, where we assume a dust-to-metal ratio of 0.4 in gas below $10^6$\,K and no dust at higher temperature due to thermal sputtering and the Small Magellanic Cloud-like dust opacity from \citet{Weingartner:2001}. The local ionization flux is updated every step a photon packet is carried out in the grid. When the photon transport is finished, we calculate the ionization state of each cell assuming ionization equilibrium. We run photon transport and update the ionization states for ten iterations to ensure convergence. Besides the $\fesc$ from the galaxy, we also track the $\fesc$ from individual star particles in each snapshot. We only consider single-star stellar population synthesis model\footnote{We use the single-star models in the Binary Population and Spectral Synthesis (BPASS) model \citep[v2.2.1;][]{Eldridge:2017}, which give consistent results to the {\sc starburst99} models.} in this paper, where almost all ionizing photons from a star particle are produced in the first 10 Myr of its lifetime. As each particle has a unique ID number in our simulations, we are able to trace particles between snapshots. With all snapshots saved in the re-simulations (0.5\,Myr between successive snapshots), we can also calculate the time-averaged $\fesc$ for all stars particles over their lifetimes. 

Using a large sample of cosmological zoom-in simulations of $z\gtrsim5$ galaxies run with the FIRE-2 model, \citet{Ma:2020} found that the sample-averaged $\fesc$ (i.e. $\fesc$ averaged over galaxies at the same stellar mass) increases with stellar mass up to $\Ms\sim10^8\,\Msun$ and decreases at the more massive end. The stellar mass of the two galaxies lie around the peak of the $\fesc$--$\Ms$ relation and they likely dominate the ionizing photon budgets toward the end of ionization (see \citealt{Ma:2020}, section 5.1).\footnote{Note that there is another simulation in \citet{Ma:2020a} that we choose not to use in this paper (z5m12b), for the following reasons. It is a massive galaxy ($\Mhalo\sim10^{12}\,\Msun$, $\Ms\sim10^{10.5}\,\Msun$) that has low $\fesc$ due to heavy dust attenuation. Also, such massive galaxies have low number densities in the Universe, so they are not the main sources for reionization. Finally, the mass resolution adopted in this simulation is $m_b\sim7000\,\Msun$, which tends to produce lower $\fesc$ compared to simulations at $m_b\lesssim900\,\Msun$ resolution \citep[see][for details]{Ma:2020}.}

\begin{figure*}
\centering
\includegraphics[width=0.85\linewidth]{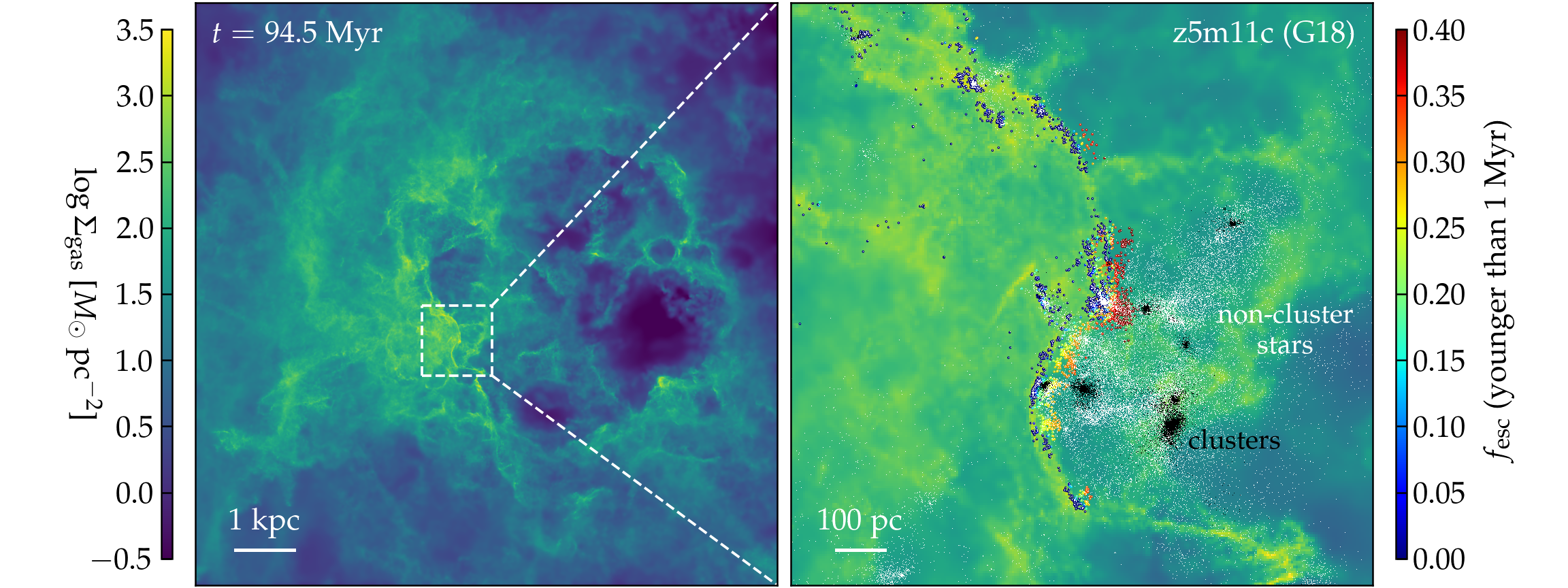}
\caption{{\em Left}: The galaxy-scale projected gas surface density for z5m11c (G18) at $t=94.5$\,Myr after the restart (10\,kpc on each dimension). {\em Right}: Zoomed image into an actively star-forming region that leaks ionizing photons efficiently, as marked by the white dashed box in the left panel (1.2\,kpc on each side). The points show all stars formed in this region in the past 10\,Myr, with non-clusters stars in white and cluster stars in black. The color points highlight stars younger than 1\,Myr, color-coded by $\fesc$. A large number of stars are formed in association with the cluster `complex' in this region. They have comparable, if not lower, $\fesc$ than the cluster stars, while they do not belong to any cluster. Moreover, stars formed in the compressed, accelerated shell at the edge of the superbubble are not bound to any cluster. They have $\fesc\gtrsim20$--40\% despite being younger than 1\,Myr. This is much higher than the average $\fesc$ from cluster stars of similar ages. On average, non-cluster stars can have as high $\fesc$ as proto-GCs. This figure is adapted from fig. 7 in \citet{Ma:2020}.}
\label{fig:shell} 
\end{figure*}

\section{Results}
\label{sec:results}
\subsection{Escape fraction from proto-GCs}
\label{sec:fesc}
To begin with, we briefly discuss the star formation history and the galaxy-averaged $\fesc$ in the re-simulated starbursts, but we refer our readers to Appendix \ref{app:sfh} (Figs. \ref{fig:fesc}--\ref{fig:tdist}) for more details. In the early stage of a starburst, the $\fesc$ from the galaxy is generally low ($\lesssim5$--10\%) because of the large amount of gas in the ISM the galaxy accumulated to trigger the starburst. The $\fesc$ increases rapidly 10--20 Myr after the dramatic increase of the star formation rate (Fig. \ref{fig:fesc}) as feedback from newly formed stars starts to clear some channels for ionizing photons to escape, which has been noted in numerous studies before \citep[e.g.][]{Paardekooper:2011,Kimm:2014,Ma:2015,Ma:2020,Trebitsch:2017,Smith:2019}. In the later stage of the starburst, the $\fesc$ maintains $\gtrsim10$--30\%. We do not find significant bias of cluster formation in the starburst, which means proto-GCs form with similar efficiencies in all stages of the starburst (Fig. \ref{fig:tdist}). One exception may be z5m10a (G18), where a considerable number of clusters formed preferentially at early time ($\sim50$\,Myr since the simulation is restarted, when the $\fesc$ from the galaxy is still low).

As the time-step between two successive snapshots (0.5\,Myr) in our re-simulations is much smaller than the lifetime a star particle produces all the ionizing photons ($\sim10$\,Myr), we can calculate the time-averaged $\fesc$ for each new star particle formed during the re-simulations using all snapshots after it formed, which we regard as the average $\fesc$ over its lifetime. Strictly speaking, this does not apply to stars formed in the last 10\,Myr in the re-simulations, but it only affects a small fraction of particles in our analysis. We classi- fy the newly formed stars in the starbursts as cluster stars and non-cluster stars, based on whether or not a particle belongs to a gravitationally bound star cluster (i.e. proto-GC) identified at the end of the re-simulation (Section \ref{sec:gc}). Fig. 2 shows the distribution functions of lifetime-averaged $\fesc$ both for cluster stars (blue) and non-cluster stars (orange) formed in each re-simulation. We do not find significant differences on the distribution of $\fesc$ between these two groups. Both cluster and non-cluster stars span a wide range of $\fesc$ from 0 to $\gtrsim0.5$. There is a higher fraction of non-cluster stars that have $\fesc\sim0$, while their $\fesc$ distribution tends to extend to higher values compared to cluster stars. We also calculate the average $\fesc$ over all stars, cluster stars only, or non-cluster stars formed in each re-simulation and list the results in Table \ref{tbl:fesc}. We do not find significant difference between cluster and non-cluster stars on their average $\fesc$ in most runs. Only in z5m10a (G18) do cluster stars show systematically lower $\fesc$ than non-cluster stars, due to the fact that a considerable fraction of cluster stars have low $\fesc\lesssim0.1$, most of which are formed in the early stage of the starburst ($\sim50$\,Myr after the restart, see Fig. \ref{fig:tdist}).

We also list the fraction of ionizing photons from bound clus- ters (proto-GCs) in Table \ref{tbl:fesc}, both for photons emitted and escaped, which we refer to as $\femcl$ and $\fesccl$, respectively. We present in Fig. \ref{fig:frac} the ratio $\fesccl/\femcl$ for each re-simulation, which is also the ratio of average $\fesc$ of cluster stars to all stars. This ratio larger than 1 means cluster stars on average have higher $\fesc$ compared to non-cluster stars. The four z5m11c runs have $\fesccl/\femcl$ within 10\% from unity, while in z5m10a (G18) clusters show 20\% lower $\fesc$ than average stars. Our results suggest that (a) proto-GCs tend to have comparable $\fesc$ to those of other stars in the galaxy and (b) the fraction of ionizing photons provided by proto-GCs for cosmic reionization is comparable to their formation efficiencies in galaxies at $z\gtrsim5$. This should apply at least to the stellar mass range we study in this paper ($\Ms\sim10^7$--$10^9\,\Msun$). 

Finally, we emphasize that the good agreement between the 4 z5m11c re-simulations suggests that our conclusion is robust to the star formation model adopted in our simulations.

\subsection{Physical insights}
\label{sec:physics}
In this section, we discuss why proto-GCs do not have significantly higher $\fesc$ than average stars in the galaxy using z5m11c (G18) as an example. In Fig. \ref{fig:fage}, we show the $\fesc$ averaged over star particles in small age bins, where we include all snapshots saved for this re-simulation in our analysis. The particles are divided into 3 groups: non-cluster stars (orange), the first half of stars formed in a cluster (thin blue), and the second half of stars formed in a cluster (thick). The first-half cluster stars show systematically lower $\fesc$ than non-cluster stars in the first 3\,Myr of their lifetime, when a star particle produces $\sim80\%$ of the ionizing photons over its entire life. This is possibly due to the fact that these clusters form at the high-density, high-pressure end of the ISM, where the optical depths are so high that ionizing photons cannot escape efficiently. As stellar feedback starts to destruct the natal clouds, the second-half cluster stars tend to have higher $\fesc$. Such an evolution trend of $\fesc$ over the lifetime of a star-forming cloud is also found in other simulations \citep{Howard:2017,Kim:2019}. On average, cluster stars do not show higher $\fesc$ than that of non-cluster stars.

In the left panel of Fig. \ref{fig:shell}, we show the gas surface density for a $10\,\kpc\times10\,\kpc$ projection in z5m11c (G18) at $t=94.5\,\Myr$ after simulation restart. In the right panel, we zoom in on an actively star-forming region of $1.2\,\kpc\times1.2\,\kpc$ as marked by the white dashed box in the left panel. The points show all the star particles younger than 10\,Myr in this region, where the white and black points representing non-cluster and cluster stars, respectively. The color points show stars less than 1\,Myr old, color-coded by their $\fesc$. It is worth emphasizing that this region is leaking ionizing photons efficiently. Note that Fig. \ref{fig:shell} is adapted from fig. 7 in \citet{Ma:2020}.

The zoomed region is at the edge of a kpc-scale superbubble. A number of clusters have formed within $\sim200$\,pc of each other in this region over the past 10\,Myr (black points). Meanwhile, a large number of stars formed spatially in association with these clusters, but are not gravitationally bound to any cluster (white points). The $\fesc$ from these non-cluster stars is comparable to that from nearby clusters. In addition, the compressed dense shell at the front of the bubble is forming stars while accelerated by feedback (likely from the clusters). Therefore, stars formed in this shell, though less than 1\,Myr old, show $\fesc\gtrsim20$--40\% (see section 4.1 in \citealt{Ma:2020} for more discussion). This is significantly higher than the $\fesc$ from average cluster stars at this age, although these stars do not form as bound clusters. For these reasons, non-cluster stars on average may have as high $\fesc$ as proto-GCs in our simulations.

Our results seem inconsistent with those from \cite{He:2020}, where the authors simulated a suite of isolated molecular clouds of various compactness and reported the $\fesc$ from the cloud increases with mean cloud density. Their work suggests that proto-GCs, presumably born in the most compact clouds, should show higher $\fesc$ than non-cluster stars born in less compact clouds. Here we briefly discuss why there is such discrepancy. First of all, \cite{He:2020} only considered photoionization feedback, while we also take into account other feedback channels including radiation pressure (non- ionizing photons absorbed by dust grains), stellar winds, and SNe. Both theories and observations suggest that radiation pressure may be crucial for disrupting the birth clouds before the first SNe going off \citep[e.g.][]{Murray:2010,Lopez:2011,Kim:2019}. In particular, \cite{Kim:2019} found high $\fesc$ both from lower-mass, low-density clouds and very compact clouds due to rapid cloud de-struction, which does not contradict our results but is not necessarily in full consistency with those in \cite{He:2020}. Next, we only consider gravitationally bound objects as proto-GCs, whereas stars that are born in the same cloud/complex but not bound to any cluster are considered as non-cluster stars. In contrast, \cite{He:2020} studied the average $\fesc$ over all stars formed in the cloud, although not all of them belong to a bound cluster. Moreover, not all stars in our simulations form in isolated clouds (e.g. young stars formed in compressed, dense shells at the front of SN-driven bubbles; Fig. \ref{fig:shell}) and many of them do not belong to proto-GCs despite having high $\fesc$. Such configurations cannot be accounted for in isolated cloud simulations. We caution that the discrepancy between the results of \cite{He:2020} and ours may not be caused by a single reason, but all factors above are likely responsible for the difference.

\section{Discussion and conclusions}
\label{sec:discussion}
In this paper, we analyze two cosmological zoom-in simulations of galaxies at $z\gtrsim5$ from the FIRE project, z5m10a and z5m11c (first presented in \citealt{Ma:2020a}), with halo mass $10^{10}$ and $10^{11}\,\Msun$ at $z=5$ (stellar mass $10^7$ and $10^9\,\Msun$), respectively. They lie around the peak of the $\fesc$--$\Ms$ relation found in a large sample of FIRE-2 simulations of galaxies at $z\gtrsim5$ and galaxies at this mass scale possibly dominate the ionizing photon budgets toward the end of reionization \citep{Ma:2020}. In \citet{Ma:2020a}, we picked up a starburst from each galaxy and re-simulated the burst with different star formation models and we saved one snapshot every 0.5\,Myr. These simulations are able to explicitly resolve the formation of compact, gravitationally bound star clusters (or proto-GCs) self-consistently. About 17--39\% of the stars formed in our re-simulated starburst belong to a bound cluster (see also Fig. \ref{fig:review} and Table \ref{tbl:resim}).

In this work, we post-process all snapshots saved for these re-simulations with a MCRT code to calculate the ionizing photon escape fraction ($\fesc$) from every star particle in the galaxy. Using the small time interval between two neighbor snapshots, we calculate the average $\fesc$ over the lifetime of every particle. We compare the $\fesc$ from proto-GCs to that from non-cluster stars to understand the contribution of ionizing photons from proto-GCs to cosmic reionization. Our main conclusions are the following:

(i) The distribution of lifetime-averaged $\fesc$ is broadly consistent between cluster stars and non-cluster stars in all re-simulations (Section \ref{sec:fesc}, Fig. \ref{fig:fdist}). The average $\fesc$ over all cluster stars is comparable to that of non-cluster stars (Fig. \ref{fig:frac} and Table \ref{tbl:fesc}). This result is robust to the star formation model in our simulations.

(ii) The first half of stars formed in any bound cluster tend to have lower $\fesc$ in the first 3\,Myr in their lives than non-cluster stars (Fig. \ref{fig:fage}), likely because clusters preferentially form in high-pressure regions of the ISM that are optically thick. The second half of stars formed in a cluster have higher $\fesc$ as feedback starts to disrupt the birth cloud.

(iii) A large number of stars form near or between clusters, or in the compressed shell at the front of the superbubble. These stars tend to have high $\fesc$, but they do not belong to any cluster (Fig. \ref{fig:shell}, Section \ref{sec:physics}). As a consequence, proto-GCs do not necessarily have higher $\fesc$ than non-cluster stars.

(iv) Proto-GCs likely contribute a fraction of all ionizing photons that reionize the Universe comparable to the cluster formation efficiency in high-redshift galaxies (17--39\% in our simulations).

One of the major uncertainties in these simulations is that the cluster formation efficiency, or the fraction of stars formed in bound clusters (see Table \ref{tbl:resim}), depends on the star formation model adopted in our simulations at a factor of a few level. Ideally, we would like to follow the formation and dynamical evolution of clusters to $z=0$ and compare with the present-day GC population. This would allow us to constrain the star formation models used in our simulations as well as calibrate the bound fraction in galaxies at $z\gtrsim5$, which is crucial for understanding the contribution of proto-GCs to reionization. However, we cannot reliably trace the dynamical evolution of clusters over cosmic time with our current resolution and N-body method (cf. section 6.1 in \citealt{Ma:2020a}). A comparison with sub-grid models of GCs in a cosmological context, where cluster forma- tion and dynamical evolution are tracked by tracer particles \citep[e.g.][]{Li:2017,Kruijssen:2019}, is worth future investigation.

Another caveat of our work is that the current mass resolution used in our simulations ($\sim100$--$900\,\Msun$) is still far from resolving the complex structure and feedback processes in molecular clouds like in isolated-cloud simulations. \cite{Ma:2020} have explored the resolution convergence of sample-averaged $\fesc$ using a suite of FIRE-2 simulations of galaxies at $z\gtrsim5$ and found comparable $\fesc$ for simulations at $\sim100$ and $900\,\Msun$ mass resolution, while those at lower resolution tend to produce systematically lower $\fesc$. They have also run extensive tests on the star formation and stellar feed- back algorithms and found no significant impact on their predicted $\fesc$ (section 5.1 therein). In this work, we also find our conclusions unchanged between z5m10a and z5m11c and among different star formation criteria considered here (Table \ref{tbl:resim}). However, we caution that our simulations do not resolve all the key physics and may not capture the correct time-scales of star formation and cloud disruption. Despite the computational challenges for resolving individual molecular clouds in great detail in current-generation galaxy-scale simulations, we emphasize the importance of pushing the numerical resolution in future work to obtain more accurate predictions of the escape fractions.

We emphasize that our definition of proto-GCs only accounts for small, gravitationally bound stellar structures. Those formed in close proximity but not bound to a star cluster will not be included when we study $\fesc$ from cluster stars. \citet{He:2020}, in contrast, reported the average $\fesc$ over all stars formed in an isolated cloud, although not all of them are in bound clusters. Moreover, our simulations resolve clusters down to $10^{4.5}\,\Msun$ in z5m11c ($10^{3.5}\,\Msun$ in z5m10a), but such low-mass clusters are unlikely to survive to $z=0$ \citep[e.g.][]{Muratov:2010}. Even more massive clusters may be destructed by tidal shocks in subsequent starbursts and mergers \citep[e.g.][]{Kruijssen:2012}. It is thus non-trivial to link the clusters in our simulations to the progenitors of present-day GCs studied in empirical models as sources for reionization \citep[see e.g.][]{Ricotti:2002,Boylan-Kolchin:2017,Boylan-Kolchin:2018}. One needs to be careful about potential differences in the definition of star clusters or proto-GCs when comparing results from different studies.

The ionizing photon leakage from the {\em Sunburst} arc come from an object with stellar mass $\sim10^7\,\Msun$ and effective radius $\lesssim20\,\pc$, which is possibly a gravitationally bound star cluster \citep[e.g.][]{Vanzella:2020}. This is comparable to the most massive cluster formed in simulation z5m11c (G18; see the cluster mass function in Fig. \ref{fig:review} and also fig. 4 in \citealt{Ma:2020a} for the formation process\footnote{An animation is available at \url{http://www.tapir.caltech.edu/~xchma/HiZFIRE/globular/Fig4_movie.mp4}. However, note that our simulations were run at $z\gtrsim5$, while the {\em Sunburst} arc is at $z=2.37$.} of this cluster). Interestingly, the {\em Sunburst} arc contains a group of smaller stellar clumps, possibly a star cluster complex similar to that in the right panel of Fig. \ref{fig:shell}. No ionizing photon leakage has been detected in this region, probably because of a low $\fesc$ or slightly older stellar ages. Also, it is worth noting that ionizing flux from compact clusters may be detected more easily than that from diffuse stars due to the high surface brightness of clusters \citep[e.g.][]{Ma:2018}. Future observations of highly magnified sources (effective spatial resolution $\lesssim10$\,pc) with compact star-forming regions like the {\em Sunburst} arc using JWST will provide better data to compare with our simulations and hence allow us to understand the nature of these objects and their ionizing photon leakage. 

\section*{Acknowledgement}
XM thanks the organizers and participants of the virtual KITP program on globular clusters in May 2020 under the unusual COVID-19 circumstances, which inspired this study. This research was supported in part by the National Science Foundation under Grant No. No. NSF PHY-1748958. The simulations and pos-processing analysis presented in this work were run on XSEDE resources under allocations TG-AST120025, TG-AST130039, TG-AST140023, TG- AST140064, TG-AST190028 and TG-AST200021.
This work was supported in part by a Simons Investigator Award from the Simons Foundation (EQ) and by NSF grant AST-1715070.
AW was supported by NASA, through ATP grant 80NSSC18K1097 and HST grants GO-14734 and AR-15057 from STScI.
CAFG was supported by NSF through grants AST-1715216 and CAREER award AST-1652522, by NASA through grant 17-ATP17-0067, by STScI through grant HST-AR-14562.001, and by a Cottrell Scholar Award from the Research Corporation for Science Advancement.
MBK acknowledges support from NSF CAREER award AST-1752913, NSF grant AST-1910346, NASA grant NNX17AG29G, and HST-AR-15006, HST-AR-15809, HST-GO-15658, HST-GO-15901, and HST-GO-15902 from the Space Telescope Science Institute, which is operated by AURA, Inc., under NASA contract NAS5-26555. 

\section*{Data Availability Statement}
The data underlying this article are available in the article and will be shared on reasonable request to the corresponding author.

\bibliography{library}

\appendix

\section{Star formation history and $\fesc$ in the re-simulations}
\label{app:sfh}
\begin{figure}
\centering
\includegraphics[width=\linewidth]{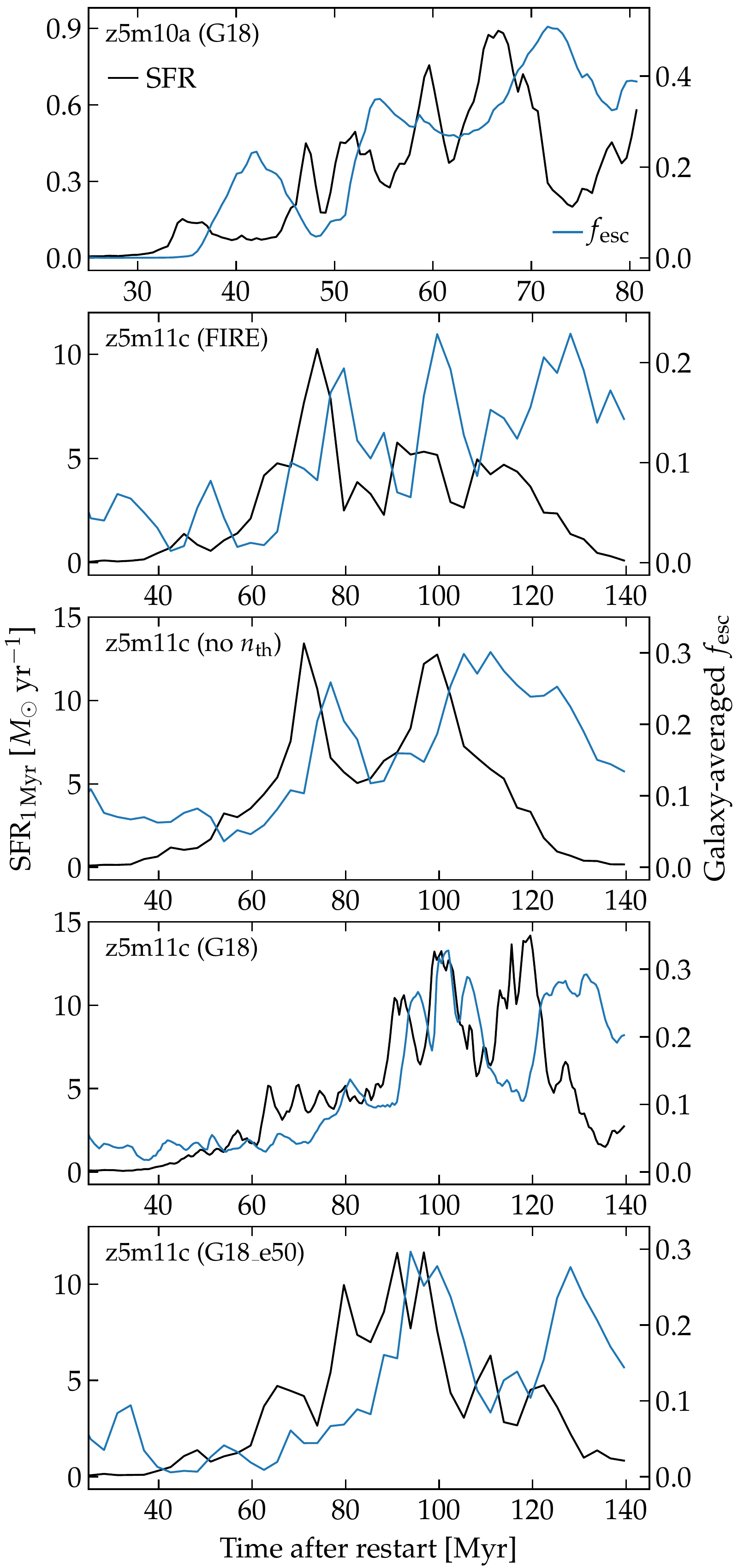}
\caption{The star formation rate (black, left) and instantaneous $\fesc$ from the galaxy (blue, right) for all re-simulations studied in this paper. The $\fesc$ is low in the early stage of the starburst, increases rapidly 10--20\,Myr since the rise of the star formation rate, and maintains $\gtrsim10$--30\% in the late stage of the starburst.}
\label{fig:fesc} 
\end{figure}

We briefly describe the star formation history and galaxy-averaged $\fesc$ in the re-simulations at the beginning of Section \ref{sec:fesc}, which we show in detail in this section. Fig. \ref{fig:fesc} shows the star formation rate (black, left) and instantaneous $\fesc$ from the galaxy (blue, right) for each re-simulation. The escape fraction $\fesc$ is low ($\lesssim5$--10\%) in the early stage of the starburst, as there is a large gas reservoir accumulated in the ISM to trigger the burst. The $\fesc$ increases $\sim10$--20\,Myr after the rapid increase of the star formation rate, as feedback starts to clear some paths for ionizing photons to escape. The $\fesc$ maintains $\gtrsim10$--30\% in the later stage of the starburst.

\begin{figure}
\centering
\includegraphics[width=\linewidth]{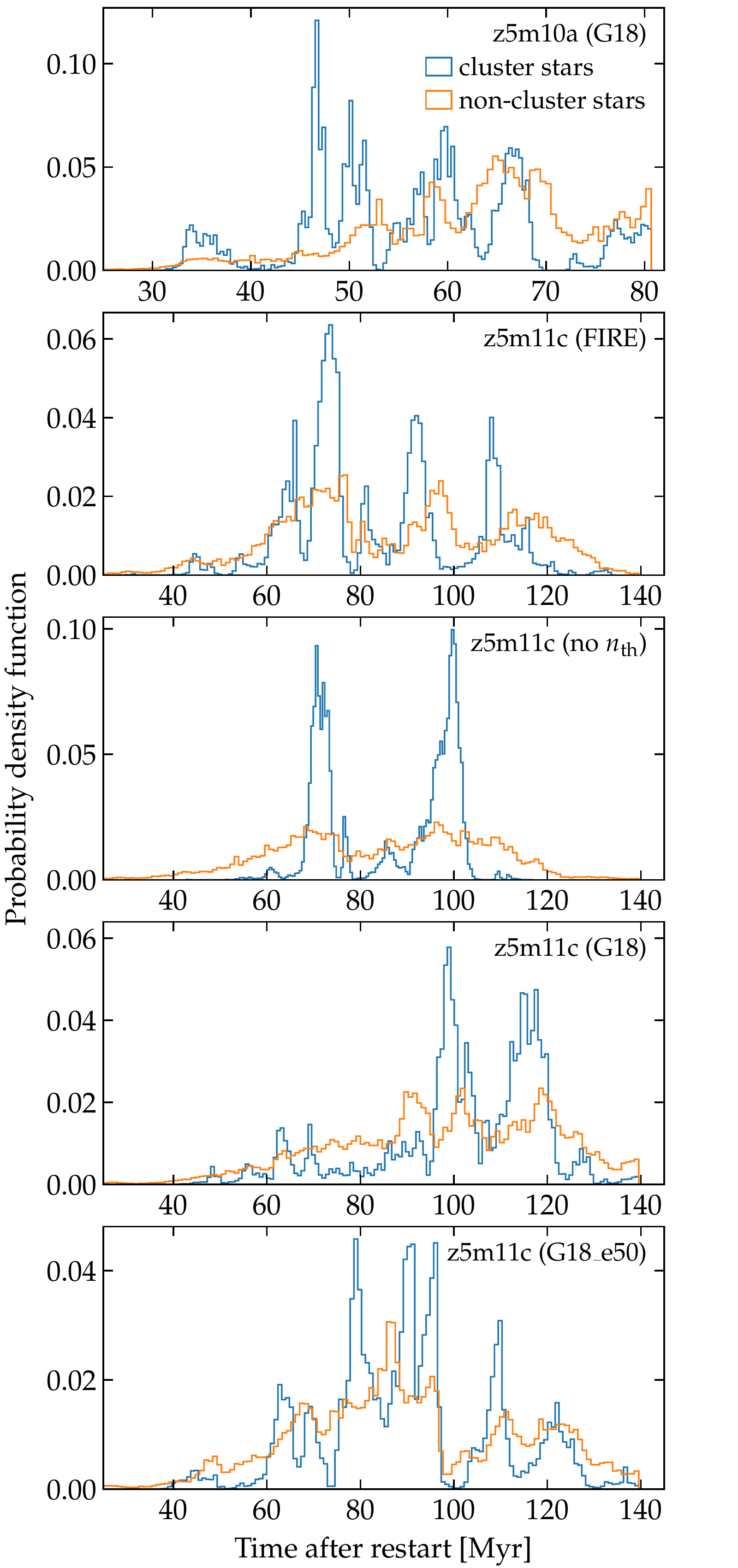}
\caption{Normalized distribution of star formation time for cluster stars (blue) and non-cluster stars (orange) in all re-simulations. Cluster formation broadly traces the star formation in the galaxy in all stages of the starburst.}
\label{fig:tdist} 
\end{figure}

Fig. \ref{fig:tdist} presents the distribution of formation time for cluster (blue) and non-cluster stars (orange) in every re-simulation. We do not see significant bias of cluster formation during the starburst. In other words, proto-GCs form at comparable efficiency in all stages of the burst. The only exception might be z5m10a (G18), in which a considerable fraction of clusters form preferentially at early time ($\sim50$\,Myr after the simulation restarted). These stars dominate the peak at $\fesc\lesssim10\%$ for cluster stars in the top panel of Fig. \ref{fig:fdist}. They also likely result in the lower $\fesc$ from clusters than that from non-cluster stars in this run (see Fig. \ref{fig:frac} and Table \ref{tbl:fesc}).

\label{lastpage}

\end{document}